\documentclass[utf8]{FrontiersinVancouver} 

\usepackage{url,hyperref,microtype} 
\usepackage[onehalfspacing]{setspace}
\usepackage{textcomp} 

\def\<{$<$}
\def\>{$>$}
\def\_{\textunderscore}

\def\keyFont{\fontsize{8}{11}\helveticabold }
\def\firstAuthorLast{Sample {et~al.}} 
\def\Authors{Giuliano Vitali\,$^{1,*}$}
\def\Address{$^{1}$DiSTAL, University of Bologna, Bologna , Italy }
\def\corrAuthor{Giuliano Vitali}

\def\corrEmail{giuliano.vitali@unibo.it}

\begin{document}
\onecolumn
\firstpage{1}

\title[Running Title]{Ontological Plant Representation for Dynamical Simulations} 

\author[\firstAuthorLast ]{\Authors} 
\address{} 
\correspondance{} 

\extraAuth{}

\maketitle
\begin{abstract}
The present study is aimed at analysing the bene?ts of an ontological approach in Functional Structural Plant Modelling. The ontological approach has been used at two levels, to re?ne the conceptual modelling approach, and to de?ne the nomenclature of the plant. To the scope available domain-speci?c ontologies describing plant entities and their relations have been analysed to verify how they support botanical and phenological descriptions at different scales. The analysis put in evidence how ontologies have a large number of shared terms and also host a large number of structural and dynamical relations among entities, however, they still lack semantic annotation useful for a complete and consistent Conceptual Modelling, as put in evidence by Foundation Ontologies. Nonetheless, the analysis also put in evidence the potential of the approach and the basis for designing a bridged ontology to be used to produce tools for learning botany, growing techniques, and supporting precision Agriculture.
\tiny
 \keyFont{ \section{Keywords:} FSPM, ontology, virtual plants, modeling, eco-physiology} 
\end{abstract}


\section{Introduction}

Functional Structural Plant Modeling (FSPM) is a discipline studying how single plants as well as vegetal tissues, populations and communities can be simulated together with their spatial representation (structure) and full interaction with the environment.
The objective is to quantitatively model how exogenous and endogenous factors affect the local environment (e.g. temperature, shading) and regulate and activate processes relevant to the plant growth and productive aspects (such as yield and quality of produce). Such a detailed representation of plants and processes can be fruitfully used for teaching purposes of botany and growing techniques, as well as in Precision Agriculture.

To simulate plant growth the most popular approach is based on L-System technique \citep{Prusinkiewicz-1990}, a geometry-oriented rule-based recursion algorithm. 

An alternative approach is given by the topological approach \citep{Honda-1971} based on dedicated data structures to store the plant state in its modules (metamers). It can be considered at the base of the Object Oriented (OO) approach, which proved to support real plant coding \citep{Vitali-2012}.

Both approaches are suggested from an apparently simple plant architecture defined on the base of naked-eye observation and on a formalism driven from the agronomic application contexts \citep{Godin-1998}.

In fact, both approaches suffer strong limitations. L-Systems hardly allow encoding of existing plants to let them to be regrown by simulation.
On the other hand, the class hierarchy of OO technique hardly allows to model the plasticity typical of living tissues (specialisation) or the appearance of new organs (e.g. adventitious buds, see e.g. \citealt{Gomez-1995}).

Moreover, though they have been used to represent plant development processes at a different scale (cell, tissue, organ), the way these processes are connected to one another can hardly be embedded in a unique computational scheme - there is still the need to have a model that incorporates processes operating at different scales.

When developing a new model of a system to be used for simulations, system engineers start from conceptual modelling characterised by three main tasks: (1) mapping to the ‘original’ system, which is expressed through a modelling language (e.g.graphical); (2) a reduction aimed at identifying a subset of the original system and (3) the pragmatics aimed at describing its purpose \citep{Verdonck-2015}.

When conceptual models are developed in a cross-disciplinary context (e.g. engineering and botany) a lack of a common agreement on terminology often occurs.
This disagreement could happen both in the graphic symbolism used in conceptual modelling, as much as in the terms used to describe the system itself.

About the first point, a standard is in use for several years, Unified Modeling Language \citep{UML-2017}, supporting several typologies of graphs to describe structural and dynamical aspects of a system.

However, UML doesn't give the modeller the rules about the elements to be drawn in the graph, as it lacks of semantics. That is the reason why in conceptual modelling, the role of Ontologies has grown considerably in the last years. In particular, Ontology-driven Model Design (ODCM) proved to be a robust and quick model development process \citep{Katasonov-2012} also useful for novice modellers \citep{Verdonck-2019}.
Reasoning over ontologies may help in discovering inferences, and locating concepts in a class hierarchy, testing the consistency of the conceptual model, and inspecting the conceptual knowledge embedded in the model \citep{Zedlitz-2014}. 

Two types of ontologies should be recognised though.
A first typology may be recognised in Foundational Ontologies, helping in classifying objects to support semantics in the modelling language \citep{Fonseca-2019}.
The second typology is represented by Domain Ontologies developed to support knowledge in a number of areas, particularly in descriptive sciences, including biology. Their terminology, however, may be so different to make them hard to be merged, and the development of a controlled vocabulary specific for plant modelling seems to be preferable \citep{SaintCast-2020}.

The aim of this work is to analyse the possibility to integrate available Domain Ontologies with the constraints of Foundation Ontology for the conceptual design of functional and structural plant modelling.
  
\section{Ontologies}
An ontology is a shared conceptualisation of a domain of interest \citep{Gruber-1993} given by terms/symbols linked by relationships with a semantic meaning.
Ontologies sinks their roots in different disciplines, including:
\begin{itemize}
    \item language theory - deriving from with logics and related to language computability, possibility to obtain a prove about truth of a statement. Ontological approach is born to be used in semantic analysis (e.g. in text mining, \citealt{Sanchez-2004}), and is currently used for bibliography research (Web Ontologies).
    \item expert systems - ontologies can be considered descendants of Expert Systems (ES), developed in the 1970's and popular in the 1990's, mainly used for diagnosis purposes and to develop decision support systems. ES host knowledge in terms of axioms, premises or facts, and inference rules, to be used to obtain conclusions (with a certainty level). Ontologies generally host a general data models, made of general types of things sharing certain properties, but they may also include information about specific individuals. The first component can be considered the skeleton of a knowledge graph, helping to validate and build new knowledge \citep{Hedden-2019, Schrader-2020}.
\end{itemize}

\subsection{Foundational Ontologies}

Despite a number of suggestions produced to drive Conceptual Modeling (e.g.Occam razor), related graphs are easily redundant and poorly consistent,and the need for a semantic represenation in UML entities has already been evidenced (e.g., \citealt{Cerans-2013}) 
A way to reduce the fuzzyness around entities has been proposed by Foundational Ontologies - here we will consider the most popular, the Unified Foundational Ontology (UFO, \cite{Guizzardi-2015}, an approach that lead to an extension of UML \citep{OntoUML-2022} introducing a constraints on entities \citep{OntoUML-2022}.

The first point in UFO is to define stereotypes, that is an object categorization. A first distinction is made between \textit{sortal} and \textit{non-sortal} types (classes of objects), the former being those endowable to some \textit{identity} (e.g.\textit{bud}) contrasting with those that cannot (e.g.\textit{tissue}).

UFO also consider the concept of \textit{rigidity}, characterising those classes that derive from a class only (e.g. \textit{vegetative-bud} $\rightarrow$ \textit{bud}) - entities in such a rigid hierarchy are called \textit{kind} (top of hierarchy) and \textit{subkind}. \textit{Anti-rigid} stereotypes are \textit{roles} and \textit{phases} (see table \ref{tab_UFO}).

\begin{table}[!htb] 
\centering
\begin{tabular}{| c | c c |} 
 \hline
  & Rigid & Anti-rigid \\
 [0.5ex] 
 \hline
      Mixin &          & Phase Mixin \\ 
            & Category & Role Mixin \\
\hline
 Sortal     & Kind     & Phase \\ 
            & SubKind  & Role \\
            & Relator  &  \\
            & Collective  &  \\
\hline
 Nonsortal  & Quantity &  \\ 
\hline
    Aspects & Mode     & \\ 
            & Quality  &  \\
 \hline
\end{tabular}
\caption{ UFO stereotypes}
\label{tab_UFO}
\end{table}

Stereotypes as \textit{phase} and \textit{roles} (referring to subjects that can be sorted) or \textit{PhaseMixin} and \textit{RoleMixin} (if they cannot be sorted) may in fact be used to define entities that can derive from different classes: let's think to the 'enlighted tissue' and 'shaded tissue', given by tissues with different 'roles' in photosynthesis during daytime, while having the same at nightime. In this example the \textit{Mixin} suffix refers to the fact that the organs (sortal) or tissues (non-sortal) can derive from different organs, as a branch or a fruit in the first case (sortal), fruit or branch skin in non-sortal case.

Another important class of sterotypes refers to relations, in which it is possible to identify some base stereotypes that interprete the aspects featuring a given type:
\begin{itemize}
    \item characterisation: a relation between a bearer entity and some features
    \item structuration: is the base constructor of any class, collecting and ascribing those features (of different typologies) characterising each class
\end{itemize}

A set of relations as
\textit{ Part-whole}, \textit{ Component\_Of}, \textit{ Containment},
 \textit{Member\_Of}, \textit{ SubCollection\_Of}, \textit{ SubQuantity\_Of}, identifies arrows commonly used in UM to connect entities. 
UFO includes relations with a logical/computational equivalent as
\begin{itemize}
    \item Formal: e.g. a vegetative bud 'is lighter than' a flower bud
    \item Material: e.g. 'flow of sap' from organ A to organ B
\end{itemize}

As OntoUML hardly supports dynamical features, other diagrams are suitable to the urpose. In particular about Discrete Event Systems quite popular  is BPMN, \cite{Rosing-2015}), that also has an ontological extention - Onto-PMN \citep{Guizzardi-2013}.
In Onto-PMN, entities \textit{participates} to \textit{events} (like previously to \textit{roles}).
Events identified as \textit{Atomic} and {Complex} (compound) are ascribed to objects (with a given role).
Most of semantic contents stay in the concept of \textit{situation} (state of affairs) that changes after the occurrence of an \textit{event} and in that of an entity \textit{disposition} (meant as power, ability, capacity,etc) that determine a \textbf{causal explainability} of the occurrence of a given event. From this viewpoint an \textit {event} determines a \textit{triggering} of a \textbf{transition}, driving a \textbf{transition rule} suitable of a probabilistic approach (causal law). It follws a distinction between 'triggering event' and 'resulting event'.

\textbf{UFO vs OO} - UFO put in evidence structures embedded in OO (UML), trying to manage them explicitly, as abstract classes, that cannot be instantiated, or some method/function that require to be written in the subclass.

Some \textit{sortal} stereotypes used to represent \textit{aspects}, as \textit{quantity} and \textit{amount}, reflect classes that can be found in OO-languages, namely Set and Magnitude (a mother classes of every numerical and measurable quantity), both characterised by large number of specialised subclasses. 
A subclass like \textbf{ordered-set} can be used to represent two UFO stereotypes as \textbf{aspects} including \textbf{mode} and \textbf{quality}.
Relations (in UML identified by arrows of several types) can also be ascribed to class ownership, often of magnitude-like or set-like: they are often associated a multiplicity.

After \cite{Tomaiouolo-2005} a major problem in OO representation of Ontologies is represented by the fact that in the last case instances are not required to be referred to some class (e.g. " the 'color white' characterizes more flowers than leaves " ).
Authors also put in evidence other main differences, tough proving that coding an ontology within an OO scheme may be complex, but also feasible.
The issue of roles has already been solved by \cite{Fowler-1997}, and it is based on definition of a specific class defining roles: ${StringCollection> roles> bud-role }$
and successively subclass every object requiring roles in a separate class branch:
${Object > ObjectWithRoles}$ including attributes to describe possible roles and the presently acting one.
The complexity of translation is also related to the language the conceptual model has to be coded. Smalltalk (supported by platforms as Squeak, Pharo, ..), as a pure OO language, is more suitable to support those paradigms, with respect to other OO-style languages (e.g. Phyton, Java, ..) which are basically based on imperative language (as C), that require a preliminary type declaration and memory allocation).

\begin{figure}[h!]
\begin{center}
\includegraphics[width=0.8\textwidth]{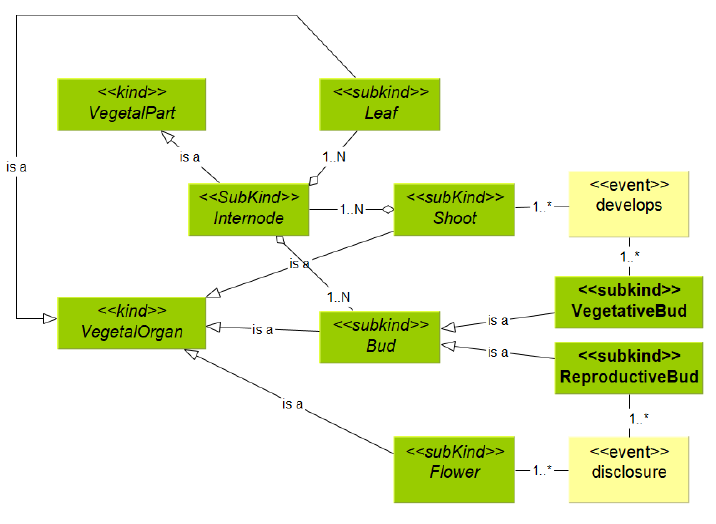}
\end{center}
\caption{Example of Onto-UML class diagram for a plant grow model}
\label{fig_1}
\end{figure}

\textbf{example 1} - In figure \ref{fig_1} an (onto-extended) Class Diagram is used to draw a OO toy model of a plant. An \textit{internode} is here used as an object that can bear a number of \textit{leaves} and \textit{buds}, the latter being specialised into \textit{vegetative-bud} and a \textit{reproductive-bud}. In a dynamical simulation framework the latter develop generating respectively shoots and a flowers, entities inheriting features from the class \textit{VegetalOrgan}, while \textit{leaves} and \textit{shoots} also beared from the \textit{internode} continue growing as any other \textit{VegetalPart}.
In the diagram they are using the suggestions of UFO, labeling entities with $\langle\langle kind \rangle\rangle$ and $\langle\langle subkind \rangle\rangle$, while the arrows are used to represent specialisation/generalisation (is a), and composition (1..N). Two processes $\langle\langle event\rangle\rangle$ are also used to transform objects (buds into flower and shoot), which in a simulation perspective, could mean destroy the first object using its parameters to create the new one (morphing).

\subsection{Domain Ontologies}
While a formal language (as UML) is represented by a reduced number of symbols (e.g.boxes and arrows to connect them), the languages characterising knowledge domains are used to generate large annotated dictionaries of terms and relations. Both, terms and relations, may be collected in a multi-hierarchical framework, and a standard is represented by OWL (Ontology Web Language) and RDF (Resource Description Framework), based on XML (Extensible Markup Language).

A growing number of tools, both on-line and desktop-based, allow to access OWL, browsing over, as \textbf{Ontobee} (operating on 251 ontologies), operate queries, as \textbf{Ontofox} (\url{https://ontofox.hegroup.org/}), or merging, as \textbf{robot} \citealt{Jackson-2019}), while for inspecting/editing \textbf{Protégé} (\url{https://protege.stanford.edu}) a widely used platform.

Ontologies can also be translated in Graphic Data Base, as in \textbf{Neo4j} (\url{https://neo4j.com/}) with the plugin \textbf{neosemantics} (\url{https://neo4j.com/labs/neosemantics/}).

The majority of ontologies can be found (and fully retreieved) at: \url{http://purl.obolibrary.org/obo/}NAME\url{.owl}.

\textbf{Ontologies for Plants} - A number of ontologies are already available and many are under development - largest ones are in the domain of medicine and biology. From some queries in Ontobee (\url{https://www.ontobee.org}) some of the largest ontologies dedicated to plant emerge, which are reported in \ref{tab_1} are together with some size indicator.

\begin{table}[!htb] 
\centering
\begin{tabular}{|c| c c c c|} 
 \hline
 NAME & content & classes & object     & annotation \\
          &         &         & properties & properties \\
 [0.5ex] 
 \hline\hline
 BTO & BRENTA Tissue Ontology & 6569 & 10 & 27 \\
 FLOPO & Flora Phenotype Ontology & 35424 & 111 & 62\\
 PO & Plant Ontology & 2018 & 300 & 190 \\ 
 PPO & Plant Phenology Ontology & 443 & 333 & 64 \\ 
 TO & Plant Trait Ontology & 5216 & 159 & 76 \\
 [1ex] 
 \hline
\end{tabular}
\caption{ List of largest ontologies characterising domains of researches on vegetal plants}
\label{tab_1}
\end{table}

 They cover a wide-range of aspects (e.g.development stages, \citealt{Pujar-2006}), but also  host annotations related to species \citep{Jaiswal-2005} useful to extend the ontology to cover particular aspects \citep{Akbar-2021}, while queries can be used to develop species-specific ontologies \citep{Walls-2019}.

\begin{figure}[h!]
\begin{center}
\includegraphics[width=0.6\textwidth]{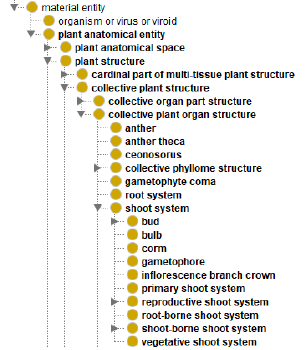}
\end{center}
\caption{Excerpt of PO hierarchy displayed in Protégé}
\label{fig_2}
\end{figure}

\textbf{Ontology structure} - Browsing Ontobee for the term \textit{bud} we get, together with its definition on PO: \textit{an undeveloped shoot system}, and its tag: \textit{PO:0000055}, other information including:
\begin{itemize}
    \item  hierarchy: \textit{Thing \> continuant \> independent continuant \> material entity \> plant anatomical entity \> plant structure \> collective plant structure \> collective plant organ structure \> shoot system}.
    In the hierarchy the root entity \textit{thing} is followed by a chain of entity types reflecting the standard adopted from ontology editing board. The \textit{material entity} is the starting point of many entities of major interest.
    
    \item subclasses: \textit{vegetative bud, axillary bud, terminal bud, reproductive bud}, are all representing specialisation of the parent entity (\textit{bud})
    
    \item siblings: \textit{root-borne shoot system, shoot-borne shoot system, primary shoot system, reproductive shoot system, inflorescence branch crown,  corm, bulb, vegetative shoot system, gametophore}, represent children of the same parent (\textit{shoot system - a collective plant organ structure that produces shoot-borne portions of meristem tissue and the plant structures that arise from them} - see figure \ref{fig_2}).
    \item relations: properties tagged from a specific Ontology (Relation Ontology - RO). Almost every of such relations are represented by directed edge on a graph, and have an inverse relation - e.g. the relation: \textit{develops\_from} RO:0002202 is the inverse of: \textit{delelops\_into} RO:0002203. Relations have a \textit{domain}, represented by those entity types they can be applied.

\end{itemize}


\textbf{Some relevant terms} - In BTO \textit{\textbf{WholePlant} - BTO:0001461 - The main part of a plant}) has a main child \textit{shoot - BTO:0001243 - a sending out of new growth or the growth sent out: as a stem or branch with its leaves and appendages especially when not yet mature}), which has the following parts:

 \begin{itemize}

     \item \textit{stem - BTO:0001300 - The main trunk of a plant; specifically: a primary plant axis that develops buds and shoots instead of roots} (with 8 subclasses);
 
     \item \textit{internode - BTO:0000636 - Region on a stem between nodes} (no descendants);
      
     \item \textit{leaf - BTO:0000713 - A lateral outgrowth from a plant stem that is typically a flattened expanded variably shaped greenish organ, constitutes a unit of the foliage, and functions primarily in food manufacture by photosynthesis}, with subclasses: \textit{ brct, leaflet, final leaf, true leaf ,etc};
      
     \item \textit{bud - BTO:000158 - A small lateral or terminal protuberance on the stem of a plant that may develop into a flower, leaf, or shoot},including in its subclasses: \textit{apical bud, dormant eye, axillary bud, leaf bud, flower bud (further specialised)}.
 
\end{itemize}

Such entities are all children of \textit{thing}.

A more detailed plant description can be found in PO, where \textit{whole plant - PO:000000} is a child of \textit{\textbf{plant structure} - PO:0009011 - A plant anatomical entity that is, or was, part of a plant, or was derived from a part of a plant}, where we may recognise two important children:

 \begin{itemize}
    \item \textit{collective plant structure - PO:0025007 - a plant structure that is a proper part of a whole plant and includes two or more adjacent plant organs or adjacent cardinal organ parts, along with any associated portions of plant tissue )}. Its main descendant is \textit{\textbf{shoot system} - PO:0009006 - A collective plant organ structure (PO:0025007) that produces shoot-borne portions of meristem tissue (PO:0009013) and the plant structures (PO:0009011) that arise from them}, whose children include:
     \begin{itemize}
     \item \textit{Bud - PO:0000055 - An undeveloped shoot system};
     \end{itemize}
 
    \item  \textit{multi-tissue plant structure - PO:0025496 - a plant structure that has as parts two or more portions of plant tissue of at least two different types and which through specific morphogenetic processes forms a single structural unit demarcated by primarily bona-fide boundaries from other structural units of different types}. Children include \textit{\textbf{plant organ} - PO:0009008 - A multi-tissue plant structure that is a functional unit, is a proper part of a whole plant, and includes portions of plant tissue of at least two different types that derive from a common developmental path}, which in turn includes:

     \begin{itemize}
     
         \item \textit{shoot axis - PO:0025029 - a plant axis that is part of a shoot system}, having as children: \textit{stem ( PO:0009047 - A shoot axis that is the primary axis of a plant}) and \textit{branch (PO:0025073, a shoot axis that develops from an axillary bud meristem or from equal divisions of a meristematic apical cell}).
         
        \item \textit{phyllome - PO:0006001 - a lateral plant organ produced by a shoot apical meristem )}, having as a childr \textit{Leaf (PO:0025034 - a phyllome that is not associated with a reproductive structure}).
         
      \end{itemize}
\end{itemize}

Every \textit{plant organs} are part of a \textit{shoot system}.

Most of these entities are used in other ontologies, where they are enriched with relations and annotations. They are also indirectly referred to in TO (Trait Ontology), e.g. \textit{branch angle  - TO:100000009 - A branch morphology trait which is the angle of a branch }) where entities support almost every phenotypical observation.

\textbf{Dynamic aspects} - In Plant Pheno Ontology (PPO) it is possible to find entities describing dynamical aspects. There is an ancestor named \textit{occurrent} having among children \textit{process} which is defined in Basic Formal Ontology as \textit{BFO:0000015}, as \textit{an occurrent that has temporal proper parts and for some time t, process s-depends\_on some material entity at t}.

\textit{Process} has a single child (in PO) \textit{\textbf{plant structure development stage} - PO:0009012 - a stage in the life of a plant structure during which the plant structure undergoes developmental processes}, with children:

\begin{itemize}

    \item \textit{whole plant development stage} 
    \item \textit{collective plant organ structure development stage} (including bud development stages);
    \item \textit{plant tissue development stage}, including development stage of vascular tissues (xylem,phloem);
    \item \textit{multi-tissue plant structure development stage}, including development stages for fruit, seed and other plant organs.

\end{itemize}

Such a hierarchical representation is different from that found in PPO, where development stages descend directly from \textit{occurent} while:\textit{biological process} include:

\begin{itemize}

    \item \textit{biological process} 
    \item \textit{collective plant organ structure development stage} (including development processes);
    \item \textit{multicellular organismal process};
    \item \textit{reproductive process};
    \item \textit{response to stimulus};
\end{itemize}

Though PO and PPO interpretation of \textit{process} is rather different to that given in NCI thesaurus - \textit{NCIT:C17828 - An activity occurring within an organism, between organisms or among organisms and the mechanisms underlying such events} (neither \textit{process} nor \textit{biological process} are used in PO, PPO or other plant-related ontologies).

\textbf{Analysing OWL} 
Looking at PO and PPO OWLs, it is possible to see the practical use of OWL language.

\textit{Class} is the main entry, used to describe and annotate terms.

\textit{AnnotationProperty} is the local dictionary of decriptors e.g. \textit{definition}, \textit{synonym}, etc.

\textit{Axioms} are assertions about a property relating a source to a target. They are used to enrich the set of annotations, and also may include supplementary definitions and synonyms (exact or narrow) using the tags defined in \textit{AnnotationProperty}.

\textit{ObjectProperty} report the relations between entities, most relevant being given by:
    \begin{itemize}
        \item
        \textit{part\_of}
        \& \textit{has\_part}
        \item
        \textit{preceded\_by}
        \& \textit{precedes}
        \item \textit{participates\_in}
        \& \textit{has\_participant}
        \item \textit{located\_in}
        \& \textit{adjacent\_to}
        \item \textit{develops\_from}
        \& \textit{develops\_to}
        \item \textit{starts}
        \& \textit{starts\_with}
        \item \textit{ends\_after}
        \& \textit{bearer\_of}
        \item \textit{generated\_from}
        \& \textit{depends\_on}
        \item \textit{developmentally\_preceded\_by}
        \& \textit{developmentally\_precedes}
    \end{itemize}

They are used (in classes) in a 'subclass-restriction', 'equivalentClass-restriction', or a 'disjointWith-restriction' context \citep{OWL-2004}.

The 'restriction' is used to define the class (as subclass, same class, or outside the reference class hierarchy) on the base of a property, which could be on the base of a parameter value or on number of allowed items ('cardinality'); here is a fragment of ontology of \textit{PO\_0001094: plant embryo coleoptilar stage}:
\textit{
        \<rdfs:subClassOf\>
            \<owl:Restriction\>
                \<owl:onProperty rdf:resource="http://purl.obolibrary.org/obo/BFO\_0000063"/\>
                \<owl:someValuesFrom rdf:resource="http://purl.obolibrary.org/obo/PO\_0001081"/\>
            \</owl:Restriction\>
        \</rdfs:subClassOf\>
},
telling that the: \textit{plant embryo coleoptilar stage}  \textit{precedes (BFO\_0000063)} the \textit{mature plant embryo stage (PO\_0001081)}.
%
%
%
%



    

Following figures have been produced importing the OWL in a GDB with Neo4J-semantics. Nodes are shown as circles with a the property \textit{label} as caption.

\begin{figure}[h!]
\begin{center}
\includegraphics[width=0.8\textwidth]{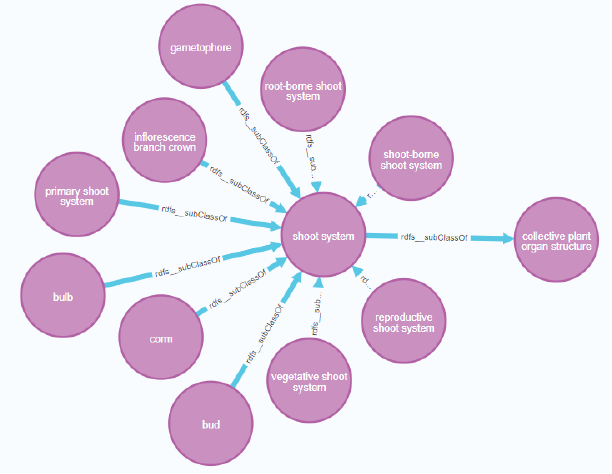}
\end{center}
\caption{Linkages among children \textbf{bud} (PO:0000055)
displayed in Neo4J Browser}
\label{fig_3}
\end{figure}

Figure \ref{fig_3} reports the set of \textit{Class} nodes centered on \textit{shoot system} from PO.

\textbf{UFO interpretation} - From an UFO viewpoint, it can be observed that entities appearing in figure \ref{fig_3} own to \textit{Kind} and \textit{Subkind} prototypes.

In those ontologies, together with organs, also appear entities as \textit{root initial cell}, and \textit{vascular System} that evidence \textit{Sortable} entities at different scales, together with \textit{Nonsortable} entities as \textit{portion of 'some tissue'}, easily related to prototypes as \textit{Quantity}.

However OWL doesn't seem to support any suggestions to identify \textit{Sortable} from \textit{Nonsortable}, not any clues about class stereotyping.

Object Properties contain features referring to space and time relative allocation of entities, easily recognised as UFO relations.
While spatial features could be enriched for detail from other ontologies (as Trait Ontology), dynamical aspects can be detailed by ontologies as PPO.

In Onto-PMN, events may include creator/destroyer of entities - in PPO family of processes includes most of GO types descending from process (including organ develpment, reproduction, dormancy, aging, response to stimulus, abscission, germination) and include both unicellular and multicellular organisms.
Such processes may easily include any metamorphoses process as well as development of new organs (latent bud).
Nonetheless, processes, as much as events, are (non material) entities to which material entities \textit{partecipates\_in} - one of the ObjectProperties listed above, e.g.
\textit{dormant leaf bud - participates in (some) - bud dormancy process}.

\begin{figure}[h!]
\begin{center}
\includegraphics[width=0.8\textwidth]{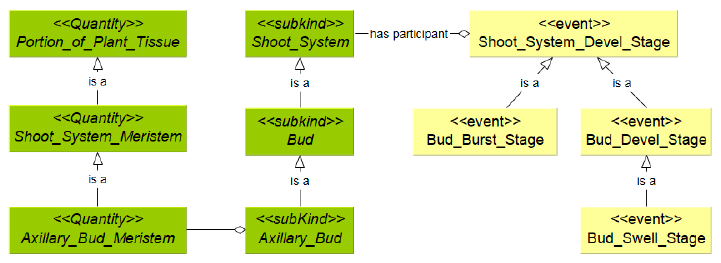}
\end{center}
\caption{Conceptual Model representing swelling of an \textbf{axillary bud}}.
\label{fig_4}
\end{figure}

Figure \ref{fig_4} reports the construction of a Conceptual Model for the dynamics of \textit{Bud} in relation to the organ bearing it \textit{Plant Shoot} and to the tissue it is made for, collecting the classes from PO (on the left). \textit{Bud} dynamics are represented by stand alone entities - in particular \textit{Bud\_Swell\_Stage} inherit feature from \textit{System\_Devel\_Stage}, has any vegetal part as a participant. This is the technique that allow in the OO paradigm to have 'messages' as standalone functions.

Dynamical aspects may be related to some change of \textit{phase}, but they hardly refer to metamorphoses of entities -  entity may disappear (or die) and be substituted by an evolved self - as both are descendant of a common class (e.g. plant part) the new entity inherit some properties, as mass and location.

Morphing, together with \textit{roles}, though implicitly present in some complementary ontologies, could require more specialised add-ons to include concepts already faced in popular branch of plant \& crop modeling (e.g.sink/source modeling), which may also include technical terms, e.g. tree growers refers to vegetal parts not in use of botanists as "Brindle", "Sucker"," Dart","Spur", etc.

\section{Conclusion}

The development of Domain Ontologies is becoming a common practice in scientific community, becoming more and more rich of descriptions about living beings' anatomy and behavior, while using terms shared by a large community.

Ontology-based Conceptual Modeling may profit of such a large amount of coded description, together with the constraints defined by Foundation Ontologies to suggest a new approach for plant structural and functional modeling.

The analysis put in evidence that ontologies offer the possibility to include in modeling a large amount of details and process occurring at several scales, and eventually to integrate them.

Domain Ontologies, however, may need a 'bridge ontology' to be merged together and enriched by semantical descriptors suggested by Foundational Ontologies. 

Because of size and complexity of ontologies, the next step should consist in the development of a tool helping in identifying missing definitions and relations, useful to improving current ontologies and make robust design Conceptual Models in the same time.

\section*{Conflict of Interest Statement}
The authors declare that the research was conducted in the absence of any commercial or financial relationships that could be construed as a potential conflict of interest.

\section*{Funding}
Not applicable.

\section*{Acknowledgments}
Not applicable.

\section*{Supplemental Data}
Not applicable.

\section*{Data Availability Statement}
The study is based on publicly available ontologies reported in the manuscript. Any detail on methodology is available directly from author.

\bibliographystyle{Frontiers-Harvard}

\bibliography{manuscript}


\section*{Figures}

\end{document}